\documentclass[journal]{IEEEtran}
\usepackage{amsmath,amsfonts}
\usepackage{algorithmic}
\usepackage{algorithm}
\usepackage{array}
\usepackage[caption=false,font=normalsize,labelfont=sf,textfont=sf]{subfig}
\usepackage{textcomp}
\usepackage{stfloats}
\usepackage{url}
\usepackage{verbatim}
\usepackage{graphicx}
\usepackage{cite}
\usepackage{graphicx}
\usepackage{cite}
\usepackage{amsmath,amssymb,amsfonts}
\usepackage{algorithmic}
\usepackage{graphicx}
\usepackage{textcomp}
\usepackage{xcolor}
\usepackage{balance}
\usepackage{booktabs}
\usepackage{float}
\usepackage{tikz}
\usetikzlibrary{matrix}
\usepackage{multirow, rotating, makecell}
\usepackage{orcidlink}

\newcommand{\sol}{\textit{QuantTM}}
\newcommand{\ie}{\textit{i.e., }}  
\newcommand{\eg}{\textit{e.g., }}  
\newcommand{\cf}{\textit{cf. }} 

\def \1{\textit{(i)}}
\def \2{\textit{(ii)}}
\def \3{\textit{(iii)}}
\def \4{\textit{(iv)}}
\def \5{\textit{(v)}}

\newcommand*{\xdash}[1][3em]{\rule[0.5ex]{#1}{0.55pt}}

\begin{document}

\title{\sol{}: Business-Centric Threat Quantification for Risk Management and Cyber Resilience}


\author{Jan von der Assen~\orcidlink{0000-0002-0591-8887},~\IEEEmembership{Graduate Student Member,~IEEE}, Muriel F. Franco~\orcidlink{0000-0002-0208-0521},~\IEEEmembership{Member,~IEEE},\\\ Muyao Dong, Burkhard Stiller~\orcidlink{0000-0002-7461-7463}}

\markboth{Preprint submitted to arXiv}%
{Shell \MakeLowercase{\textit{et al.}}: A Sample Article Using IEEEtran.cls for IEEE Journals}


\maketitle

\begin{abstract}
Threat modeling has emerged as a key process for understanding relevant threats within businesses. However, understanding the importance of threat events is rarely driven by the business incorporating the system. Furthermore, prioritization of threat events often occurs based on abstract and qualitative scoring. While such scores enable prioritization, they do not allow the results to be easily interpreted by decision-makers. This can hinder downstream activities, such as discussing security investments and a security control’s economic applicability. This article introduces \sol{}, an approach that incorporates views from operational and strategic business representatives to collect threat information during the threat modeling process to measure potential financial loss incurred by a specific threat event. It empowers the analysis of threats’ impacts and the applicability of security controls, thus supporting the threat analysis and prioritization from an economic perspective. \sol{} comprises an overarching process for data collection and aggregation and a method for business impact analysis. The performance and feasibility of the \sol{} approach are demonstrated in a real-world case study conducted in a Swiss SME to analyze the impacts of threats and economic benefits of security controls. Secondly, it is shown that employing business impact analysis is feasible and that the supporting prototype exhibits great usability.

\end{abstract}


\section{Introduction}
A threat model is an abstraction of a valuable system that comprises potential security damages to be inflicted by a malicious actor~\cite{coretm}. An accurate threat model is key for technical (\eg network operators and cybersecurity analysts) and business teams to make decisions on how to protect their systems. To create such a model, the companies' systems need to be modeled, and threats must be identified, prioritized, communicated, and countered by countermeasures. Threats vary from human to device-centric, such as exploiting potential attack vectors in network devices, Web services, Internet-of-Things (IoT), or even humans that have a key role~\cite{shostack}.

Threat modeling is a system-centric activity that is affected by the technical complexity of the system, relevant threats, and their countermeasures~\cite{samm}. Threat prioritization is critical to developing strategies for an adequate cybersecurity investment and the mitigation of threats. This prioritization is vital, especially because it is impossible to mitigate all threats due to the fact that the number of threats is usually large, and countermeasures are never perfectly efficient. Therefore, efficient threat analysis and strategies are needed. Also, it is important to understand that there is no perfect security (\ie zero risk) and that resources (\eg money, personnel, and time) to implement security strategies are limited~\cite{shostack,threatmodelingmanifesto}.

From the business perspective, it is even more relevant to prioritize threats accurately~\cite{rcvar}. After all, the set of identified countermeasures must be effective while minimizing the overall cost, including the cost of the financial impacts due to cyberattacks and deployment of security control. For example, even though a threat may be technically relevant to a system, it should not be prioritized if the potential loss is negligible or if it is in contrast with the cost of the security control. Aside from being economically relevant, to enable meaningful communication of a threat model with the business side, the representation of a threat's importance metric must speak the language of businesses' decision-makers (\ie managers). The decision-makers may not understand a qualitative severity score -- however, they may understand a threat's potential financial impact.

Different limitations are frequently observed when existing threat modeling tools and methods are employed~\cite{coretm, threatmodelingmanifesto}. It can be observed that \1  threat prioritization is either not considered, \2 the resulting threat prioritization is based on a qualitative importance metric, and \3 there is a clear lack of prioritization techniques based on financial loss. Within the realm of risk management, quantification has already been explored from a variety of angles~\cite{oppliger}. However, actual deployments and successful experiences are still not widely recognized, especially when comparing the management of communication services to services from other areas, such as insurance and banking. Furthermore, experiences from risk management are not directly transferable to threat modeling since risk management spans a wide range of activities (\eg asset management and vulnerability management). Thus, during the design stage of a service, methods such as risk-based vulnerability management (\ie an implementation concern) are not conceptually aligned with the stage in the life cycle. Here, threat modeling serves as a mechanism to shift the focus to security earlier~\cite{threatmodelingmanifesto}.

Therefore, there exists a clear opportunity to apply economic models and methods to the management and prioritization of threats. Economic models have been successfully applied to plan long-term cybersecurity investments by relying on benchmarking methods~\cite{cybersececonomics, CyberTEA-Paper, gordon-loeb}. However, there is a lack of applications of economic models for systematic threat modeling and threat prioritization in today's communication systems and their underlying infrastructure. This is in stark contrast to the current understanding of resilient IT service continuity management, which mandates not just a business continuity plan, but also to identify the impact of service disruptions in a business-driven manner. Thus, Business Impact Analysis (BIA), a model that identifies and assesses the effects of business interruptions, can serve as an ally to bring a business-oriented threat analysis to the service design step, where the functional assets (\eg network and software architecture, business processes) and data assets (\eg patents, user data, analytics) are defined -- revealing not only the potential attacks that may apply to the service but also align the efforts on the ones most impactful to the business side.

To overcome this gap, this paper introduces \sol{}, a practical approach for determining the impact of threats on businesses and prioritizing decisions based on economic aspects. \sol{} defines a three-layered methodology to gather the input variables needed to quantify a threat, which includes the views and expertise of representatives from the information and communication systems, the business process, and the strategic perspectives. Once the values are collected on all three layers of an organization, \sol{} implements a practical quantitative model to relate all threat effects and their organization-wide impact for each threat, thus, producing a quantitative metric related to the financial loss. The feasibility of the \sol{} approach is showcased in a case study conducted with a real-world Swiss SME, where \sol{} was demonstrated to support all key steps needed for effective threat prioritization and security control assessment. Furthermore, these results are contrasted with existing methods. The key contributions of this article are:
\begin{enumerate}
    \item[\textit{\textbf{i)}}]
    The analysis of current threat modeling tools and methods based on literature review and vendor inquiry.
    \item[\textit{\textbf{ii)}}] An overarching methodology that incorporates threat modeling on the technical level; threat interpretation on the procedural level; and business impact analysis on the organizational layer.
    \item[\textit{\textbf{iii)}}] A business continuity model that aggregates the views from these distinct layers, allowing threat prioritization and control cost discussion on the technical layer.
    \item[\textit{\textbf{iv)}}] An approach for business impact analysis that adopts the semantics of threat modeling and the prototype to execute the approach in a visually guided and automated manner. The prototype comprises several baseline impacts identified from the literature.
    \item[\textit{\textbf{v)}}] The deployment of \sol{} as part of a case study in a real Swiss SME, enabling the discussion of the real-world applicability of business-driven threat analysis.
    \item[\textit{\textbf{vi)}}] A focus group-based experiment and illustrative, expert-based examinations discussing and contrasting the usefulness, functionality, and limitations of the BIA method and the developed prototype.

\end{enumerate}

The remainder of this paper is structured as follows. Section \ref{bg} reviews key concepts of threat prioritization and related work. Section \ref{sol} introduces the overarching \sol{} approach, which combines views from the technical and business perspectives. Section \ref{bia} further proposes a threat modeling-oriented business impact analysis process and a related set of dimensions, which fits into the \sol{} approach that is covered by a prototypical implementation. Section \ref{eval} contains the evaluations of the approach. Finally, Section \ref{conc} concludes the paper and highlights future work.

\section{Background and Related Work}
\label{bg}
This work focuses on prioritization within the threat modeling context. Explicitly, following the terminology from the National Institute of Standards and Technology (NIST), a \textit{threat} is a potential attack initiated by a \textit{threat actor} exploiting a system flaw (\ie \textit{vulnerability}) which leads to an adverse impact~\cite{nistriskassessmentguide}. While threat models focus on the first two dimensions (\ie threats and their actors) to identify, discuss, and prevent structural vulnerabilities, risk management tries to approximate all parameters. To understand the gap and possible synergies between the two practices (\ie threat modeling and risk management), related work in both areas is surveyed.

\subsection{Threat Modeling Approaches}
In order to appropriately prioritize threats and their respective countermeasures, an importance metric for the set of elicited threats must be derived. Thus, a method to assess the specific importance of a threat is needed. Depending on the method, the result differs in its scale, which provides critical context. There appears to exist no commonly agreed definition of which dimension of a threat (\eg exploitability, damage, or mitigation cost) should serve as a dimension when prioritizing. Popular methods like \textit{DREAD} or \textit{PASTA} qualitatively combine multiple dimensions to propose a scoring system~\cite{survey}.

Aside from the chosen dimension, the methodology applied to derive the importance value is critical, since it defines the measurement scales on which such a prioritization is expressed. Without knowing the measurement scale, the interpretation is skewed. The simplest measurement scale is the nominal scale, which merely expresses categories without order or distance (\eg \textit{confidentiality}, \textit{integrity}, or \textit{availability}). The ordinal scale is more complex since it expresses order among items without expressing distance between them (\eg \textit{low}, \textit{medium}, and \textit{high}). A notion of distance among candidates is expressed on the interval scale. However, only the ratio scale allows the expression of priority with purposeful proportions. To enable this comparison, a meaningful notion of a true zero point must be present~\cite{scales}. Thus, it is critical to understand not only a priority score but the underlying scale. 

With this background in mind, the remainder of this section reviews methods from the industry and from academia. For each method, the prioritization metric (\ie the qualitative or quantitative value to rank elements) is discussed along with the threat prioritization method (\ie how metrics are established).
Although there are only few methods like DREAD which are specifically dedicated to threat prioritization, many threat modeling methods and tools from the industry and academia include a threat prioritization step. Except for \textit{LINDDUN} and \textit{STRIDE}, all threat modeling methods include this prioritization step. However, some of them like PASTA do not explicitly define how a threat should be prioritized, although they include risk-based notions of assets and threats. Other methods like \textit{DREAD} explicitly define, in a qualitative way, how a set of dimensions can be assessed by assigning a score from 0 to 10. Finally, an overall score from 1 to 50 is obtained and mapped to labels from \textit{low} to \textit{critical}. Thus, the resulting score should be interpreted on the ordinal or interval scale.

\setlength{\tabcolsep}{5pt}

\begin{table}[h]
\centering
\caption{Overview of Threat Modeling Tools and Methods}
\begin{tabular}{@{}lllll@{}}
\toprule
& \textit{Work}                     & \textit{{Threat Prioritization  }} & \textit{Importance Metric}  \\ \midrule
 \parbox[t]{2mm}{\multirow{8}{*}{\rotatebox[origin=c]{90}{\textsc{Methods}}}}
 
& \textit{STRIDE}\cite{stride}           & None                & None                   \\ 
& \textit{LINDDUN}\cite{linddun}         & None                & None                   \\
& \textit{DREAD}\cite{dread}             & DREAD               & Score \textit{(I)}       \\ 
& \textit{CVSS}\cite{tmmethods}          & CVSS                & Score \textit{(I)}       \\
& \textit{PASTA}\cite{pasta}             & PASTA               & Score \textit{(O) }       \\ 
& \textit{PbD}\cite{jesus}               & DREAD-based         & Score \textit{(R)}          \\ 
& Attack Centric \cite{attackcentric}                   & Custom              & Score \textit{(R) }         \\ 
 \parbox[t]{2mm}{\multirow{13}{*}{\rotatebox[origin=c]{90}{\textsc{Tools}}}}
& \textit{Security Cards}\cite{tmmethods}& None    & None                   \\  \hline
& \textit{OWASP TD}\cite{coretm}         & Severity Estimation & Categories \textit{(O)}  \\ 
& \textit{CAIRIS}\cite{coretm}           & Risk-based & Categories \textit{(O)}           \\ 
& \textit{TRIKE}\cite{coretm}            & Exposure Estimation & Score \textit{(I)}      \\ 
& \textit{MTMT}\cite{mtmt}               & Priority Estimation & Label \textit{(O)}       \\ 
& \textit{ThreatModeler}\cite{threatm}       & Severity Estimation & Label \textit{(O)}       \\
& \textit{SeaMonster}                    & None                   & None \\ 
& \textit{IriusRisk}\cite{coretm}        & No Response & Categories \textit{(O)}  \\ 
& \textit{SDElements}                    & \multicolumn{2}{c}{\xdash[3em] No Vendor Response \xdash[3em]} \\ 
&\textit{securiCAD}                    & \multicolumn{2}{c}{\xdash[3em] No Vendor Response \xdash[3em]} \\ 
& \textit{Tutamantic}                    & Actuarial models, & Categories (O) \\ 
&                                       & Markov-chain  & Numerical Score \\ 
& \textit{Threagile}\cite{coretm}        & Risk-based & Labels \textit{(N)} \\ 
& \textit{ThreatSpec}\cite{coretm}       & None & Manual Labels \\ \hline
&   \sol{} (This work)                              & Business-centric     & Financial Loss \textit{(R)} \\ \hline
\end{tabular}
\\\textit{O} ~= Ordinal, \textit{N} ~= Nominal, \textit{I} ~= Interval, \textit{R} ~= Ratio

\label{tab:toolsandmethodstable}
\end{table}

The analysis of the documentation of industry tools (\eg \textit{CAIRIS} and \textit{TRIKE}) shows that different dimensions are considered, such as risk-based estimations, exposure estimations, or severity assessments. Also, most of the tools make use of a scoring or labeling-based result that is best interpreted on an ordinal scale. Table~\ref{tab:toolsandmethodstable} presents a mapping of industry methods and tools associated with vendors that replied to our contact during the development of \sol. For example, OWASP Threat Dragon, a popular open-source tool, is used by allowing the user to manually rate the severity of a threat on a scale of \textit{low}, \textit{medium}, or \textit{high}. However, the tool does not directly support how these values are established or interpreted.

Since many industry tools are closed-source, it is not always feasible to assess the prioritization methodology from publicly available information. After all, it has to be decided whether a numerical score holds the notion of a true zero point to be classified as a ratio-scaled score. Therefore, numerous companies were inquired by the authors of this work about the methodology that is employed. At the time of writing, only one vendor replied, stating that \textit{Tutamantic} leverages standard statistical analysis based on actuarial models and Markov-chain analysis. Risk is inferred from the number and complexity of inbound flows (\eg ports and endpoints) for each component in the system architecture. \textit{Tutamantic} provides a quantitative rating of threats based on a statistical analysis involving actuarial models and Markov-chain analysis, which is aggregated into three categories. Downstream components can leverage the numerical representation.

Overall, there is no commonly agreed importance metric and methodology to assess the importance of threats. It is important to highlight that most approaches either exclude threat prioritization or use a scoring system on an ordinal or interval scale, making the comparison of threats difficult. Finally, the lack of solutions employing economics-based approaches is evident, which are helpful in communicating the business-level impact of a threat and, thus, increasing the collaboration between technical and business stakeholders. In general, most threat modeling approaches do not enable direct translation of the results of this technical activity. While still fruitful in the technical system engineering context, data collection and the interpretation of the resulting prioritization for non-technical collaborators remains a key challenge.

\subsection{Risk Management Approaches}
Given that there is a lack of readily available quantification in threat modeling, quantification approaches from the cyber risk management field are explored. Table~\ref{tab:riskmethods} presents an overview of selected work for risk prioritization and cyber risk quantification. These approaches were selected due to their maturity, the techniques applied and scenarios covered, and their relevance to the scope of this work. For each work, the goal and the approach chosen are elicited. Also, based on experiences shared by the authors and on theoretic reasoning of input parameters, the applicability to the threat modeling domain is discussed.

\begin{table*}[t]
\centering
\caption{Risk Management Prioritization and Quantification Approaches}
\label{tab:riskmethods}
\begin{tabular}{@{}lllll@{}}
\toprule    
\textit{Work}  & \textit{Year} &  \textit{Goal}  & \textit{Approach} & \textit{Applicability}
\\ \midrule 
\cite{cvar} & 2022 & Quantifying Exposure & Value-at-Risk & Cyber Insurance \\
\cite{risk-insurance} & 2022 & Quantifying Exposure & Dynamic Algorithm & Cyber Insurance \\
\cite{risk-estimation} & 2022 & Rarity Estimation & Qualitative Questionnaire & Senior Security Managers \\
\cite{bayesian-fair} & 2020  & Loss Prediction & Bayesian Network & Generated data \\
\cite{maritime} & 2023  & Risk Identification & Bayesian Network & Maritime Systems \\
\cite{calcrisk} & 2020  & Maturity Benchmarking & Qualit. Questionnaire & IT leaders of SME \\
\cite{costestimation} & 2002  & Downtime Cost Estimate & Formula & System Modelling \\

\bottomrule
\end{tabular}
\end{table*} 

\textit{Value-at-Risk} is widely used in the finance world. Due to the fact that in cybersecurity, the application of the model is not widely adopted, \cite{cvar} presents a system to calculate cyber value-at-risk. Thus, the goal is not to identify relevant threats but to assess the overall risk exposure of an existing information system. By means of a case study conducted in the insurance field, the applicability of the model was demonstrated. However, from the threat modeling perspective, the model relies on parameters that are not conceptually aligned, such as the defense probability of an existing security control. Within the threat modeling context, such controls may not exist yet since the system may still be in a design stage, where it is only possible to reason about structural vulnerabilities. Similarly, the authors report the criticality of datasets being available to apply the approach, which may not be aligned with such a design activity.

Similarly, \cite{risk-insurance} has been proposed and evaluated in the insurance context. Again, conceptual differences between the two domains become apparent. For example, the algorithm proposed takes as input the degree of vulnerability to a threat, which may not be applicable in a software design phase, where arguably system flaws may not be manifested and only exist in a structural way (\ie being influenced by architectural choices). In that line, the work is comparable to~\cite{bayesian-fair}, which also relies on the parameter. Nevertheless, certain findings from~\cite{risk-insurance} are relevant to this problem and should be considered even in the threat modeling field. One of the key findings of their work is the temporal dependency of a quantification, which should be considered when creating a threat model, too.

Opposing risk quantification, \cite{risk-estimation} argues that in-depth quantification of risks is not realistic in practice since most practitioners rely on simple metrics to understand the annualized effects of threats. To understand the possibility of being affected by rare threats, a qualitative, questionnaire-based tool is proposed that surveys the perceived possibility of a breach. The tool consolidates views on ongoing breaches, which are not conceptually aligned with a system design activity. Nonetheless, the paper highlights important practical findings, such as the bias towards highly common threats. Furthermore, the authors validate the inclusion of diverse stakeholders for ad-hoc estimations and highlight that the risk management field suffers from overly complex implementations. A similar line was followed by \cite{maritime}, who applied a rule-based analysis using Bayesian networks, where parameters were modeled using linguistic terms (\ie qualitative). Aside from the literature, expert opinions were integrated into the risk assessment~\cite{maritime}, similar to the online questionnaires implemented by~\cite{calcrisk}, which was used to systematically benchmark the maturity of SMEs' technical security postures.

In summary, quantification is still not widely employed in the risk management practice for computer networks and communication services, especially compared to other fields such as banking and finance. Nevertheless, there are more risk management approaches proposed than in the threat modeling field since these two areas show conceptual differences. Therefore, risk management approaches cannot be directly applied to threat modeling, thus, showing the need and opportunity for the proposal of specific and practical approaches like \sol. Another interesting observation is the opposition of industry practitioners \cite{risk-estimation}, which highlights the complexity of existing tools and argues that a full threat-and-vulnerability mapping is not practical \cite{oppliger}. In favor of practical, expedient, and ad hoc analysis, simple approaches like~\cite{costestimation} embedded in a technical and organizational context can be used for a specific goal-oriented application.

\section{The \sol{} Approach}
\label{sol}
\sol{} addresses the challenge of interpreting threat models within a business context. The \sol{} approach provides a methodology and the accompanying terminology to translate and prioritize threats from both technical and business perspectives. A quantitative practical model is also provided as part of the approach to measure potential financial loss incurred by specific threat events.

\begin{figure}[b]
    \centering
    \includegraphics[width=\linewidth]{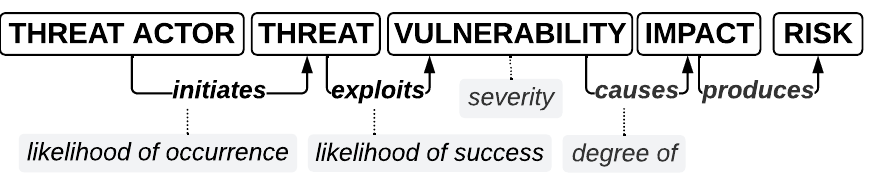}
    \caption{Key Factors in the NIST Risk Model~\cite{nistriskassessmentguide}}
    \label{fig:nist-risk-model}
\end{figure}
\subsection{Methodology}
Although the term \textit{threat modeling} is often used as a synonym for risk assessment~\cite{oreilly}, there are clear differences between the two. Threat modeling is usually considered to be a design activity within the secure system design life cycle~\cite{oreilly, wheeler}. Even within this context, one can observe an overlap between risk assessment and threat modeling. For example, NIST defines a set of key factors that comprise a risk model, as shown in Figure~\ref{fig:nist-risk-model}. In contrast, threat modeling aims to establish potential threats and their sources. Therefore, as threat models are abstractions, an abstract notion of the threats' impact must be considered during threat quantification.

\begin{figure*}[t]
    \centering
    \includegraphics[width=.9\textwidth]{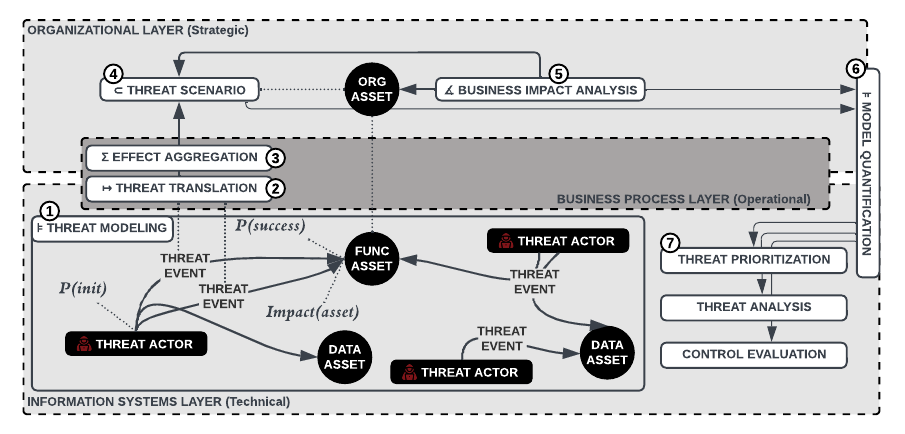}
    \caption{Three-layered Threat Quantification Methodology Indicating the Steps (White Boxes) and the Resulting Artifacts (Black Elements)}
    \label{fig:arch}
\end{figure*}

Figure \ref{fig:arch} shows the three-layered methodology defined by \sol{} for the threat quantification. The flows of the methodology are defined as follows. The threat model (Step 1) is defined on the \textit{Information Systems} layer, translated (Steps 2 and 3) and interpreted (Step 4) on the \textit{Business Process} layer, and the impact is assessed (Step 5) on the \textit{Strategic} layer of an organization. Finally, the variables resulting from applying the methodology are used as input variables for the quantitative model (Step 6) for the economic measurement. Downstream activities (Step 7) then make use of the output (\eg prioritization and control evaluation). The details of each layer that composes the methodology are provided in the rest of this section. Furthermore, the quantitative model is introduced as a key and final part of the \sol{} approach.

\subsubsection{Information Systems Layer}
The first step in the methodology defines the threat modeling (step 1) on the Information Systems (IS) level. For that, a technical perspective shall drive the identification of threat actors, their potential \textit{threat events}, and the \textit{target assets}. In enterprise risk management, where many organizations define their strategic "crown jewels," this asset identification can consider the strategic definition of crucial assets. Besides recording the discrete variables for threat events (\ie the set of threats $T$) and assets (\ie the set of assets $A$), the probabilistic variables $P_i$ and $P_s$ should gather an estimation of how likely such attacks are to be executed and how likely they are to succeed. Aside from estimates, empirical approaches can provide additional data sources~\cite{rcvar}. Depending on the type of attack, the optional value $d$ records how long a threat event will impact the organization. For many events, this variable would be $d_n = \infty$, since it will impact the asset forever. For example, a data leakage attack would forever compromise the data's confidentiality. In another direction, a DDoS attack may only lead to an impact regarding data availability for a certain time window.

Once a threat model $TM$ is created on the technical level, there exists a relation between the set of threats and their assets, noting for each relation the previously discussed probabilities (\cf Equation~\ref{eq:threatmodel}). The threat model has to be translated from the system perspective to the Business Process (BP) layer. 

\begin{equation}\label{eq:threatmodel}
\begin{split}
T = \{ t_1, t_2, ..., t_n \} \\
A = \{ a_1, a_2, ..., a_n \} \\
TM = T \times A 
\end{split}
\end{equation}

\subsubsection{Business Process Layer}
This layer entails the translation (Step 2) of each threat into one or more threat effects and the translation of an asset to the business asset supported by the technical asset. A simple approach for translation can be followed by discussing each threat event and how it affects the Confidentiality, Integrity, Availability, and Accountability (CIAA)~\cite{wheeler} dimensions of the business function. An existing definition of business assets from a previously conducted risk assessment can serve as an anchor to translate the technical asset. This step aims to define a strategic formulation of the threats' effects for each technical threat. For example, a Ransomware attack elicited on the IS layer can be translated into leading to temporary data unavailability. With the rise of extortion threats, the same event may lead to a permanent loss of data confidentiality.

A first reduction of the threat model is appropriate at this step. The involvement of the next layer (\ie organizational) should be motivated by the threat events being likely to occur at some point in time (even when this may rarely be the case) and leading to significant impacts on the business process with a specific business goal (\ie business mission). Thus, besides defining the set of threat effects (Step 3) related to each threat, the \textit{degree of impact} provides an optional variable $deg$ to hold the compromise ratio on the asset. For example, both a DDoS attack and a Ransomware attack could lead to a successful disruption of data availability. However, a DDoS attack may only lead to external unavailability -- internal sales employees may still be able to access the data. 
In a hypothetical scenario where 20\% of sales revenue is generated by back-office employees, the impact can be described with the effect of "data unavailability." Taking into consideration the business process, the impact can be discounted to 80\% since it affects only external sales employees. Concerning the likelihood, it is critical to include reasonably likely threats but not to exclude rarely occurring threats. The key is to include threats that will occur at least once over the following years (\ie excluding highly hypothetical threats).


This step requires close collaboration between business mission representatives and a technical representative who was involved during the threat model's creation on the IS level. The resulting output from this step is that for each threat, with its previously assessed likelihood parameters, a set of threat effects ($E$) are associated (\cf Equation~\ref{eq:threateffects}). This comprises an aggregated set of threat scenarios ($TS$) that can be discussed on the strategic level to capture the organization-wide impact of the scenario.

\begin{equation}\label{eq:threateffects}
\begin{split}
E = \{ e_1, e_2, ..., e_n \}\\
TS = T \times E
\end{split}
\end{equation}

\subsubsection{Organizational Layer}
Once threat events with their respective likelihoods, organizational assets, and the degree of impact of the events are mapped to threat scenarios (Step 4), the business impact of these scenarios can be assessed. This cannot be entirely established on the BP layer, as a series of intangible factors must be considered. For example, the previously defined impact on sales numbers could be approximated by a process manager. 
However, intangible factors, such as damages to publicity, relations, employee morale,  or opportunity costs, require an overarching view. For that, research on business process analysis has presented solutions to quantify the cost of a threat effect~\cite{quant-bia}. Furthermore, recently emerged cyber risk management practices have established the usefulness of relying on empirical data to approximate impacts to the organization~\cite{rcvar, SecRiskAI-Paper-CBI}. As such, \sol{} is agnostic to the business impact assessment method, given that it produces a useful cost measure as an actual number for each threat effect as shown in Equation~\ref{eq:bia}. 

\begin{equation}\label{eq:bia}
\begin{split}
C = \{ c_1, c_2, ..., c_n \}\\
BIA = TS \times C
\end{split}
\end{equation}

Ideally, intangible costs should be included in this metric, especially for strategically important assets. Section~\ref{bia} introduces a novel method to achieve such an analysis, highlighting the elicitation of tangible and intangible factors. At this point, the BIA can be defined as a set of cost factors related to each threat scenario. The quantitative model can be evaluated once the cost factors are established since all relevant input dimensions are defined.

\subsection{Quantitative Model}
\label{sec:model}
The quantitative model is applied once the cost factors have been established (Step 5) and attached to threat scenarios that are linked to threat events. These events can be quantified (Step 6) by producing a discounted cost (\ie scaled by probability and degree) metric using Equation~\ref{eq:quanttm}. For illustration, cost factors $C$ are replaced with the single loss expectancy $L$ of the threat effect given the threat event's duration.

\begin{equation}\label{eq:quanttm}
\begin{split}
Q(t) = \sum_{e_1}^{e_n} P_i(t) \times P_s(t) \times L(e_i, d=\infty, deg=1.0) 
\end{split}
\end{equation}

It is critical that all variables are explicitly recorded and communicated back to the IS layer since threat modeling is a design activity to identify, evaluate, and implement security controls (Step 7) that optimize overall cost. Based on that, existing economic metrics (\eg ROSI and Gordon-Loeb) \cite{cybersececonomics} can be evaluated during the system's design by offsetting implementation and threat effect costs of different security controls. Having all information accessible on the IS and BP layers is critical to support additional discussions with the strategic management layer to mitigate the impacts of threats using the cybersecurity budget available. However, some threats must be motivated by creating awareness by strategic management. For example, the threat effect of seeing data encrypted could be mitigated on the information systems level by implementing a read-only offsite backup strategy. However, negotiating a DDoS protection plan with an upstream Internet Service Provider (ISP) provider may involve strategic awareness by the organizational management of the dependency, as well as additional budget to implement the control. Thus, to address these different requirements and scenarios, the proposed quantitative model provides explicit numeric statements (\ie quantification) on how the threats' impact can be assessed. 

\section{Business Impact Analysis}\label{bia}
In the \sol{} approach, the BIA is positioned as a central element without specifying how organizations should implement such analyses. Nevertheless, as BIA is central to the approach and not well explored in the threat analysis context, it is necessary to develop a specific BIA method that connects to semantics from the threat modeling and translation steps.

BIA takes a business perspective on disruptions in a critical business service or mission. Aside from identifying potentially disruptive events (\ie threats, as established during threat modeling), impacts to various aspects need to be considered, thereby including a broad range of factors, such as financial, legal, or technical ones. In practice, two challenges need to be addressed: \textit{(a)} identify the effects of threat scenarios on the business service, and \textit{(b)} measure or estimate those effects using metrics that cover the aforementioned domains. 

To achieve this analysis, a set of steps must be performed to make up the proposed BIA process and lead to cost quantification. The process and the steps involved are highlighted below, followed by details of the aggregating model and the prototype guiding this process in an automated manner.

\subsection{Process}
\begin{figure}[t]
    \centering
    \includegraphics[width=\linewidth]{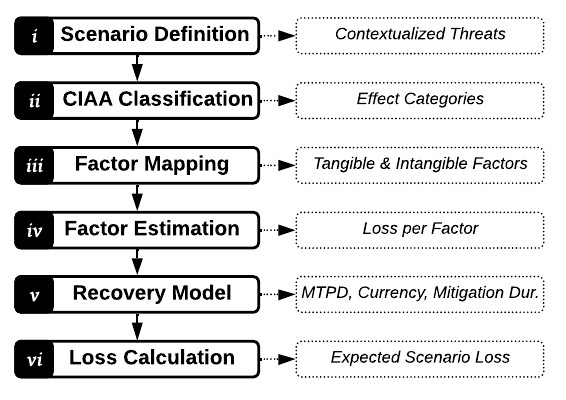}
    \caption{Steps and Resulting Artifacts Involved in the BIA Process}
    \label{fig:bia-arch}
\end{figure}

A process composed of six steps is proposed to perform BIA for a given threat. \figurename~\ref{fig:bia-arch} outlines the guiding steps of the BIA method, which require at the very beginning a previously identified threat from the technical layer of the threat model. In the first step, the \textit{(i)} scenario definition requires that the threat is considered within its business mission context. For example, the DDoS attack would be contextualized by the information system targeted and the business mission that is supported (\eg a DDoS attack on an e-commerce ordering web service). To move closer to the goal of analyzing that threat from a business perspective, the threat scenario is \textit{(ii)} classified based on relevance to the CIAA security objectives (\ie whether it affects the confidentiality, integrity, availability, and accountability) of the involved functional or data asset. Using these effect categories, various factors that \textit{(iii)} map to the objectives are derived. For \sol{}, 16 different business impact factors are collected from industry reports such as~\cite{mossburg2016beneath} and~\cite{corallo2020cybersecurity}. For example, while the defined tangible factors include direct product loss during a disruption or the violation of commercial customer agreements, intangible ones include degraded customer relationships. The complete set of factors can be found in the publicly available prototype~\cite{sourcecode}.

At this step, these factors suggest to the analysts which factors to model, although the BIA method is flexible to integrate additional ones. In the impact library, both tangible and intangible factors are provided based on their relevance to the security principle. For example, for the aforementioned availability attack, the process would suggest covering the effect of \textit{revenue loss}, but not the impact of a \textit{customer breach notification}, since data may not be leaked from this specific threat scenario. It is important to note that the presented effects are merely a guide; users can incorporate additional effects that might be particular to their business, industry, or use case. Next, the \textit{(iv)} loss estimation is performed per factor. Here, the BIA method presents persistent and one-time loss functions depending on the type of threat. 

After the critical business impacts are identified and estimated, the \textit{(v)} attack recovery and related information, such as currency and maximum tolerable period of loss (MTPD), are collected to provide the \textit{(vi)} loss calculation, as last step, for each threat, which may involve multiple threat scenarios that each lead to multiple business impacts. As part of the last step, further information might be required. For example, for persistent impacts (\ie time-dependent ones), their daily loss and likely duration are aggregated. To improve the accuracy of the loss trend, the recovery level, defined as the ratio of the given situation compared to a normal one, is also required. The model defined in in Equation~\ref{eq:bia-model} implements the relation of threat scenarios to costs defined earlier in Equation~\ref{eq:bia}, although it does not factor in the likelihood of the events from a technical perspective.


\begin{equation}
\begin{split}
    \text{Loss} & = \sum_{i} \left( \sum_{t} \text{BI\_pers}_{i} \times (1 - \text{rec}_{t}) \times \text{days}_{t} \right) \\ 
 &+ \sum_{j} \text{BI\_ot}_{j} \label{biCal}
\end{split}
\end{equation}\label{eq:bia-model}

The total loss of each threat consists of two parts, one-time loss and persistent loss. \texttt{i} and \texttt{j} itemize each persistent impact and one-time impact, and \texttt{t} represents the recovery stages of each persistent impact. \texttt{BI\_pers} represents the input loss of persistent loss, \texttt{rec\_t} is the recovered level of the business with time going on, and \texttt{days\_t} represents the time when the business restores to \texttt{rec\_t}. The approach then adds all persistent losses, which vary at different times, and all one-time losses together.

\subsection{Prototype Implementation}\label{protoype}
A prototype of \sol{} is implemented to allow users to perform a business impact analysis of a threat scenario. This enables a guided analysis while automatically collecting all relevant user input. The collected data can then be used to automatically compute the aggregated loss of impacts and to visually break down the factors involved, thereby generating insights on how business continuity could be ensured against the threats under analysis.

To minimize data privacy concerns, the \sol{} prototype operates entirely client-side without any third-party persistence components involved. The Web-based user interface relies on a set of plain HTML files with embedded functionality implemented in JavaScript. No dependencies except the browser-level \textit{localStorage} persistence API are required to operate the tool, with simple hyperlinks interconnecting the otherwise static pages in the application.

To start the BIA procedure, users enter a list of threat scenarios in the \textit{home page} of the tool. This page then heuristically attempts to provide an initial mapping to the CIAA principles that later guide the selection of impact factors. The CIAA classification is performed by matching against a set of more than 20 commonly observed threats, which can be extended by manual input. Thus, tuples of threats and a set of CIAA properties are persisted, and the user is in series redirected to a page representing the CIAA properties (\eg \textit{confidentiality.html}, \textit{integrity.html}). Here, tangible and intangible factors are suggested for the threat, which can then be activated for each of the properties. Furthermore, manual ones can be integrated. For example, for a threat that leads to a scenario impacting \textit{data availability}, the following tangible business impact factors are suggested:
\begin{itemize}
    \item Product revenues loss during the duration of service disruption
    \item Violation of commercial agreements with customers 
    \item Regulatory penalties
    \item Quality degradation of products
    \item Technical investigation cost
    \item Defense improvements (incident response, post-mortem analysis, mitigation)
\end{itemize}
Furthermore, a set of intangible impact factors is proposed:
\begin{itemize}
    \item Insurance premium increase
    \item Lost future contract revenue
    \item Customer relationship degradation
\end{itemize}

\begin{figure}
    \centering
    \includegraphics[width=\linewidth]{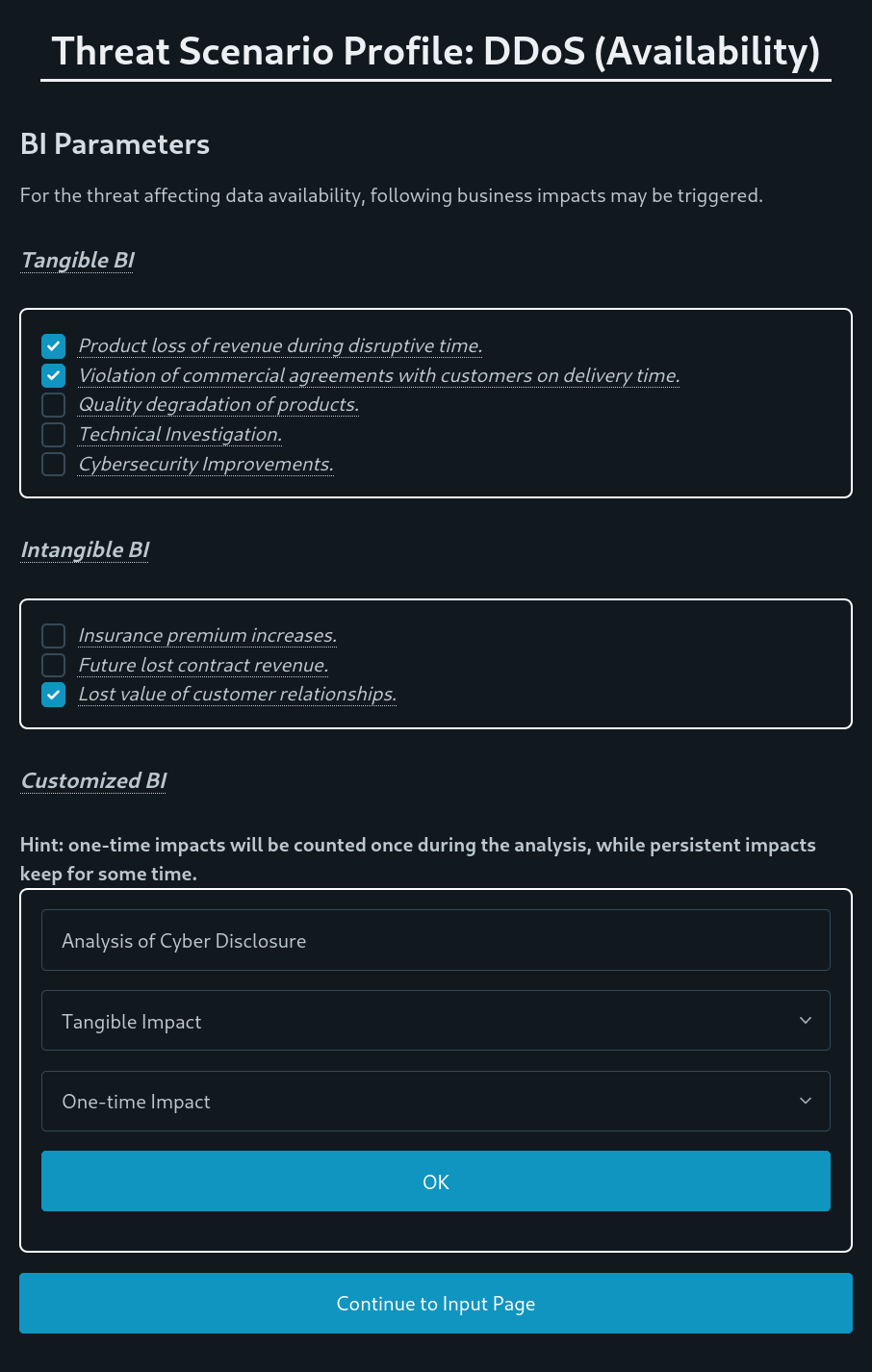}
    \caption{Impact Factor Identification for an Attack on Service Availability}
    \label{fig:bia-prototype-screenshot}
\end{figure}

For each of the business impacts, the application considers them as either one-time impacts or persistent impacts. For example, a regulatory penalty may be applied once, whereas product revenue loss depends on the outage duration. After collecting the set of applicable factors (\cf in \figurename~\ref{fig:bia-prototype-screenshot}), the \textit{input page} guides the collection of losses related to the business impacts. Thus, for one-time impact factors, each factor is displayed, and input is collected. For persistent impact factors, the recovery of the business impact can be modeled by providing tuples of recovery level and time frame as well as the estimated financial impact. In cases where the recovery function should be modeled trivially, users can simply define tuples with extreme values. For example, the following tuple would model the total outage over four business days.
\begin{equation}
(\text{recovery}=0, \text{timeframe}=4)
\end{equation}
After collecting impacts and recovery functions, the users are redirected back to the \textit{home page}, where the overall impact is computed and the results are visualized. Based on Equation~\ref{eq:bia-model}, the total loss of each threat is computed. Using bar charts, the impacts of different threats can be compared. Furthermore, for each threat, it is explained how the total loss is derived. First, pie charts display the overall distribution of tangible and intangible loss, making up the total loss. To better understand these factors, the distribution of losses among the impacts related to the threat is visualized. This is critical, as it may help business owners mitigate threats even from a non-technical perspective. For example, it may be the case that an unlikely threat leads to high impacts due to a \textit{violation of commercial agreements with customers}. Now, if all technical measures are exhausted, but the threat scenario is still unacceptable, business owners could consider re-negotiating the commercial agreement. Finally, from a resilience and business continuity perspective, the different recovery levels can be visualized by plotting the persistent losses of impacts, showing how different impacts evolve over time, allowing the discussion of how the systems or processes could be made more resilient and potentially smooth the curve.


\section{Evaluations}
\label{eval}


From the collective experience of practitioners, qualitative requirements are available for the threat modeling process. OWASP~\cite{owasp} states that a threat model should help justify security efforts by formulating assurance arguments that defend the security, thereby making the spending, efforts, and developments of the information system accountable. The working group behind~\cite{threatmodelingmanifesto} defined several critical dimensions and effects of a threat model. Specifically, a method shall be \textit{systematic} while including multiple \textit{information sources}. They must consist of \textit{varied viewpoints}, leverage \textit{useful tools}, and most importantly, they should enable translation from the theoretical model into a \textit{practical benefit}. Ultimately, threat models must be feasible and provide value to the involved stakeholders. 

Therefore, \sol{} is assessed based on these aforementioned properties. For that, a case study from a real-world deployment of the method is provided, and a focus group-based experiment of the business impact analysis procedure and tool is conducted. Finally, two expert-based experiments are conducted based on which the presented approach is discussed.

\subsection{Real-world Case Study}\label{sec:casestudy}
To assess the feasibility and value-adding properties of the \sol{} approach, the evaluation method was applied in a real-world Swiss SME in the situation of implementing an e-commerce solution over a purely manual order-by-phone process. The main goal of the experiment is not to produce an exhaustive list of threats and vulnerabilities but to develop a set of recommendations that \1 are relevant, \2 understandable for the business, and \3 that represent opportunities for actions and improvements. The considered e-commerce solution represents an exchange of an existing information system (\ie using electronic mail and telephone to order) with a Web-based information system. 

\begin{table}[h]
\centering
\caption{Company Involved in the Case Study}
\label{tab:company}
\begin{tabular}{@{}llllll@{}}
\toprule    
            \textit{Information} & \textit{Value} 

            \\ \midrule
           Sector & Trade (B2B, Machinery) \\
           Size & 10–20 Employees (SME)\\
           Country & Switzerland (Production, Logistics), USA, Spain \\
           Information Systems & ERP, CRM, Storage \& Domain Services\\
           \bottomrule
\end{tabular}%
\end{table}

The company involved in the case study can be characterized with the dimensions shown in Table~\ref{tab:company}. Within Switzerland, SMEs make up 99.72\% of the economy in terms of the number of companies while employing 66.90\% of the workforce ~\cite{sme}. Such companies are affected by cybersecurity issues particular to these types of enterprises (\eg lack of dedicated security personnel, IT budget limitations)~\cite{smeissues}. Although the selected case appears as a small company, within the Swiss economy, where an SME usually considers less than 250 employees, only 4.56\% of SMEs engage more than 20 employees, highlighting that the enterprise depicts a realistic scenario~\cite{sme}.
The company was approached via e-mail, agreeing to participate in the threat modeling workshop and its accompanying business impact analysis. To effectively apply the approach, the threat modeling workshop, conducted on the IS layer, was facilitated by the authors to introduce the method. Direct involvement with the organizational side was not needed following the translation of the threat events. Therefore, interactions via e-mail were used to request the impact analysis, supplying the key details.

\subsubsection{Information Systems Threat Model}

Initially, a threat model was created with respect to the technical specification of the information system at hand. This was created as defined in the methodologies' IS layer. As the threat universe can never be fully enumerated, a base set of threats was considered for evaluation. Thus, on the IS layer, technical representatives agreed on five key threats for the Web-based system under analysis, as shown in Table~\ref{table:threatmodel}. In addition to the threat events, the probabilities were estimated. For \textit{DDoS attacks}, it was estimated that there is an attack every five years (\ie a yearly probability of $P_i=0.2$) and that attacks are always successful (\ie an attack success of $P_s=1.0$). Here, the estimation was based on the service providers' management expertise, but it can also be based on external reports~\cite{rcvar}. 

\setlength{\tabcolsep}{5pt}

\begin{table}[h]{%
\caption{Information System Threat Model}
\label{table:threatmodel}
\begin{tabular}{@{}llllll@{}}
\toprule    
            \textit{Threat Event}   & \textit{$P_i$} &  \textit{$P_s$} &  \textit{Duration} & \textit{Degree} & \textit{Effects}
            \\ \midrule
            DDoS & 0.2 & 1.0 & 48 & 1.0 & Service Unavailability\\
                 &     &     &    &     & (External)\\

           \midrule
            CSRF & 0.4 & 0.1 & $\infty$ & 1.0 & Malicious Input \\
           \midrule
            XSS & 0.4 & 0.5 & $\infty$ & 1.0 & Malicious Input \\
           \midrule
            XXE & 0.4 & 0.1 & $\infty$ & 1.0 & Malicious Input \\
           \midrule
            Deserialization & 0.2 & 0.2 & 72 & 1.0 & Lateral Attacks, File \\
                            &     &     &    & 1.0 & Deletion, Unavailability \\

           \midrule
            Ransomware & 0.6 & 0.2 & 72 & 1.0 & Unavailability \\
                       &     &     &    &     & (Int., Ext.), Data Leak\\

           \bottomrule
\end{tabular}%
}
\end{table}
Based on the estimations backed by industry reports, 48 hours was considered a pessimistic but realistic duration for the threat event. Furthermore, since the information system is operated by an SME, the event was rated as leading to a full impact on availability. The subsequent three threats were modeled as residual threats with a low likelihood of initiation based on the expertise of the service provider and the system administrators. For example, \textit{XML External Entity} threats were considered with a low likelihood of success since XML was not used to represent messages. \textit{Vulnerable deserialization} was estimated with a higher likelihood of success because they could involve more attack paths.

The collaborative discussion and perceptions of the SME's technical team on potential threat effects (on the technical level) were critical for this step. The threat event may lead to remote code execution, which was modeled as a set of derivative threat effects with a duration of 72 hours. Finally, ransomware was analyzed as a potential threat event since the company was already targeted by such an attack, even before the ordering process was digitalized. For this analysis, up-to-date information on threat actors' behaviors is vital. The main threat event used to be data availability. However, recent years have shown an adapted behavior, where the derivative threat effects frequently include data leakage with subsequent extortion that involves companies being threatened to pay a ransom not to have sensitive data revealed to competitors.

At this point, the technical threats relevant to the information system at hand were established. Each threat event has a specific technical asset, including functional assets (\eg the shop's availability to service orders and employee's ability to process orders) and data assets (\eg confidentiality of orders and customer information). A likelihood estimation, degree of impact, and duration were determined for each. Furthermore, each threat event included one or more threat effects, which were still expressed with respect to the information system.

\subsubsection{Business Process Threat Model}
Based on the technical descriptions of the threat model, threat effects can then be translated and aggregated from a technical perspective into the effects on the business process layer. The goal of this procedure was to capture a statement representing the threats' effects on the business process without including any technical details. This step required the closest degree of collaboration. Thus, a business service representative and the system administrator of the SME have a vital role in these steps. As highlighted in Table~\ref{table:processmodel}, the probabilities are not communicated to the managers to prevent biases towards rare but high-impact scenarios.

\setlength{\tabcolsep}{6pt}

\begin{table}[h]{
\caption{Business Process Threat Model}
\label{table:processmodel}
\begin{tabular}{@{}llllll@{}}
\toprule    
            \textit{Threat Event}   & \textit{Duration} & \textit{Degree} & \textit{Business Disruption}
            \\ \midrule
            DDoS &  48 & 1.0 & \1 Customers can not order via  \\
             &   &  & shop, only via phone \\
           \midrule
            CSRF &  $\infty$ & 1.0 &\1  Malicious orders to existing  \\
             &  & & or new customers \\
           \midrule
            XSS &  $\infty$ & 1.0 &\1  Malicious orders to existing  \\
             &  & & or new customers \\
           \midrule
            XXE &  $\infty$ & 1.0 &\1  Malicious Orders to existing \\
             &  & & or new customers \\
           \midrule
            Deserialization  & 72 & 1.0 &\1  Employees cannot process \\
              &  &  & existing orders \\
            &    & &\2  Customers cannot order via  \\
            &    & &  shop, only via phone  \\
           \midrule
            Ransomware & 72 & 1.0 & \1 Employees cannot process \\
             &  &  & existing orders \\
            &    & &\2  Customers cannot order via  \\
            &    & &  shop, only via phone  \\
             &    & & \3  Existing orders are   \\
             &    & & publicly accessible  \\
           \bottomrule
\end{tabular}
}
\end{table}

Finally, several business process threat effects/impacts were identified in the SME. Some of the threats could already be analyzed with respect to their direct financial loss to the business process. However, this would ignore the details of the SME's business model. Although the SME derives a substantial part of their turnover from sales via the shop, much of the customer income is generated through other business channels (\eg consulting services and training). Therefore, it was critical to understand that the business operates on a Business-to-Business (B2B) model, with long-lasting customer relationships and high trust. Furthermore, only a few players are servicing the specialized market segment. The customers use the products sold for their core businesses -- any fault cases quickly lead to losses and endanger customer retention.

\subsubsection{Organizational Threat Model and Interpretation}
\label{interpretation}

The organizational threat model captures the effects of a technical threat on a business function for the entire organization. Thus, a management representative was asked to perform the final step of assessing the business impact of the previously defined business function disruptions. As such, the impact was quantified based on estimations (\eg assumed reaction of customers on a trust breach) and historical data (\eg impact on financial gains made in previous years). Table~\ref{table:bia} shows the results of the business impact analysis conducted based on the description of the disruption and its duration. All numbers have been collected in Swiss Francs (CHF) and converted to US Dollars (USD) using the currency rate of March 26, 2023 (\ie 1.08 USD for 1 CHF)

\setlength{\tabcolsep}{6pt}
\begin{table}[h]{
\caption{Business Impact Analysis}
\label{table:bia}
\begin{tabular}{@{}llrlll@{}}
\toprule    
            \textit{Threat Event}   & \textit{Business Disruption} &  \textit{Impact (USD)}
            \\ \midrule
            DDoS &   \1 Customers can not order via shop, & 1,620   \\
             &      only via Phone & \\
           \midrule
            CSRF  &\1  Malicious orders to existing or & 0  \\
             &    new customers \\
           \midrule
            XSS  &\1  Malicious orders to existing or & 0 \\
             &    new customers \\
           \midrule
            XXE  &\1  Malicious Orders to existing or & 0 \\
             &    new customers \\
           \midrule
            Deserialization  &\1  Employees cannot process orders &  3,240 \\
            & \2 Customers cannot order via shop  & 1,620 \\
           \midrule
            Ransomware  & \1 Employees cannot process orders  & 3,240 \\
             &    \2 Customers cannot order via shop & 1,620 \\
             &    \3 Orders are publicly accessible & 1,080,000 \\
             &    to both customers and competitors & \\
           \bottomrule
\end{tabular}
}
\end{table}

The first threat event was quantified with a direct loss impact of 1620 USD, since customers would simply order via the phone, but costs to restore the functionality would incur. The following three threats were considered unpractical from the business perspective since, in their business model, the company establishes contact first when onboarding new customers. The threat events related to \textit{Deserialization} consider first the problem of orders not being processed. In the worst case, they would have to pay for expedited delivery fees to react to customer complaints, which are assessed at 3,240 USD. Finally, in the ransomware case, the data leakage and extortion threat must be highlighted. Based on the business model, every customer and every competitor (\ie the upstream manufacturer) has specific purchasing conditions. In the best-case scenario, this would decrease profit margins since they must adapt prices for certain customers. In the worst case, they would lose their upstream procurement conditions to other competitors, costing hundreds of thousands of US Dollars since some customers would be lost in the long term. This threat could be existential to the business's long-term success.

With this analysis, the \sol{} model can be applied to quantify each threat by establishing the overall impact of one technical impact based on the impact analysis conducted. Table~\ref{table:quantm} highlights the final values of $Q(t)$ (\ie the final loss metric discounted by probability and degree) for each threat identified in the first step (IS level). It is important to note that most values are only valid for the stated threat duration. Therefore, applying the formula represents the last step in the case study conducted with \sol{} in the company.

\setlength{\tabcolsep}{10pt}

\begin{table}[h]
\centering
\caption{Quantified Threat Model}
\label{table:quantm}
\begin{tabular}{@{}llr@{}}
\toprule    
            \textit{Threat Event}  &   \textit{Threat Duration}  & $Q(t)$ (USD)
            \\ \midrule
            (Distributed) Denial-of-Service & 48 &  324\\
            CSRF & $\infty$ &  0\\
            XSS & $\infty$ &  0\\
            XXE & $\infty$ &  0 \\
            Deserialization & 72 & 194\\
            Ransomware & 72 &  13,543\\
           \bottomrule
\end{tabular}
\end{table}

Based on these results obtained in the real-world case study, several observations are drawn. First, as indicated by the quantified model, one can see that, based on the business view, three threats can be considered with low priority. The reasoning is that, based on the business-centric view captured by the \sol{} methodology, non-technical controls on the process layer (\ie personal controls of new customers) make it less likely that the threat leads to a significant impact. Secondly, two threats (\ie \textit{DDoS} and \textit{Deserialization}) can lead to significant impacts. The benefits of a quantitative approach allow the application of related models from the cybersecurity economics field. For example, the ROSI, defined as the difference between the cost of the mitigated impact and the cost of the security control, can be applied. Considering a hypothetical DDoS prevention service that costs USD 540 per year, the ROSI might yield negative returns if the mitigation rate is not good enough. Thus, the security control is not cost-effective, especially since it will unlikely reduce the attack success factor (\ie the model parameter $P_s$) to zero. Such metrics can also be leveraged for prioritization. With that, threats can be ranked by impact metric to understand the most impactful threats or by their potential return on investment to understand the opportunities.

Finally, the application of the \sol{} approach highlighted that among the threats considered, one threat (\ie \textit{Ransomware}) event is existential. This quantification can serve for different discussions. For example, on the technical level, additional security controls or even alternative system designs can be evaluated. The threat could be mitigated on the process layer by adapting the business process that spans the different systems. Here, existing business continuity methods can be applied. For instance, mitigation through transparent pricing policies could decrease profit margins; however, it could also increase customer loyalty, decreasing the threat of customers leaving altogether. Finally, on the organizational level, additional business missions could be considered -- since this threat exemplifies the firm reliance on this business function and, thus, a strategic imbalance in this income stream.

\subsubsection{Discussion}
A quantitative threat modeling approach based on business continuity factors can act as an enabler for a number of value-adding activities, such as security control cost analysis, threat prioritization, or threat mitigation on various levels of an organization. Based on the experiences from employing \sol, it is clear that the proposed approach presents different advantages. 
It was also demonstrated that \sol{} is a feasible approach even for small companies with limited technical expertise or without previous knowledge or experience in business impact analysis or business continuity management.
The \sol{} approach achieves the goal considering different \textit{information sources}~\cite{threatmodelingmanifesto} since it relies on information from different \textit{viewpoints} of an organization. This helps to identify critical threats that have a particular impact on an organization. Furthermore, it allows the identification, evaluation, and discussion of mission-critical threats. In that way, it allows for formulating \textit{assurance arguments} in the language of the three organization levels. 

Also, relying on a clearly defined methodology and a quantitative model, the \sol{} approach is \textit{systematic}. For example, the \textit{Ransomware} threat was deemed existential and communicated as such due to the preservation of the business meaning. This criticality stems from the data extortion threat being subsumed. Interestingly, this threat effect is novel since, some years ago, the threat actors' main goal was to make data unavailable. 
As highlighted by the declarative factorization of threats, the quantification is explicit and repeatable. Another important observation can be drawn when the outcome is compared to other approaches. For example, in a risk matrix (\cf Section~\ref{sec:comparison}), such an improbable but highly impactful threat would just be rated as \textit{medium}. However, as discussed in the previous section, the threat scenario could lead to a disaster, impacting the company's existence. While it would not be fair to claim that one is more correct than the other, the result from \sol{} could be considered more straightforward to understand since specific influencing factors can be discussed. This is critical in light of "long-tail" risks, which are hard to prioritize using quantitative risk management approaches and thus can only be mitigated using resilience concepts~\cite{hunziker}.   

Connecting technical notions of a threat model with its business meaning can be highly useful in the context of the \textit{defense in depth} model. As shown by the application of \sol, the business process controls can substantially influence the outcome of a threat event. For example, in the discussion of threat effects, data leakage was found to be a disaster due to how the customer relationships were structured. This highlighted not only that threats can be mitigated using technical and organizational controls but also that this sensitivity was not known to the engineers.
Nevertheless, this also introduces a set of challenges, such as the quantification approach applied here relies on the presence of a business representative. 
Furthermore, specific development models (\eg open-source and governmental efforts) do not allow an apparent involvement of a financially motivated business actor. Similarly, the evaluation based on a case study can only provide limited evidence of the method's feasibility. Nevertheless, the application within an SME indicates feasibility even with limited in-house expertise and resources.

\subsection{Focus Group Experiments}
While the previous section assessed observations from deploying \sol{} in a real-world service definition scenario, focus group experiments were conducted to understand, more specifically, how well users can leverage the method, model, and prototype. Here, particular emphasis was placed on the business impact analysis to understand how well an existing threat model can be analyzed. Focus groups are a qualitative research method involving several subject matter experts to gain insights based on personal experiences regarding the subject hand~\cite{powell1996methodology}. Here, special emphasis was placed on the business impact analysis method and prototype usage, not on the threat modeling stage.

\subsubsection{Methodology}\label{sec:focusgroup-methodology}
The focus group discussion was designed for a small group of 4-5 attendees, leveraging online conferencing as a means of communication. The subject, referred to as the evaluation target, is the implemented prototype, which implicitly embeds the analysis method. Participant recruiting was concluded without requiring participants to perform any previous work (\eg familiarizing with threat analysis or business impact analysis). Before the discussion, a quick overview of the topic and framework is given. Here, \1 the workflow and design of the proposed framework were given, \2 background knowledge was introduced to cover the core concepts involved in the framework, \3 the agenda of the discussion was introduced, including \4 the specific tasks that participants were asked to solve and \4 distribution of the questionnaire that captures views based on the System Usability Scale (SUS) as well as open feedback forms.

After introducing the framework, the actual discussion involved presenting a hypothetical case study comprising scenario description, task definition, and data elaboration. The participants were required to conduct a threat analysis for a hypothetical US-based technology manufacturer, with three threat scenarios under consideration: malware affecting the central server of their core product, an insider threat modifying customer data, and a botnet shutting down the production line. Furthermore, the participants received a description of company data from which impacts could be inferred. A set of tasks had to be performed for each of the threats. First, participants had to interpret the threat regarding the business process building from a threat scenario. Secondly, for each threat scenario, the participants had to leverage the prototype to identify potentially relevant business impact factors. Finally, participants had to quantify the impacts, which were captured by answering questions directly corresponding to functional and qualitative requirements of the main functions of the prototype. The questions are defined as follows.

\begin{quote}
\begin{enumerate}[]
    \item[\textbf{Q1}] When choosing the business impacts, do you understand what it evaluates?
    \item[\textbf{Q2}] Which threat costs the most to address, and which is most emergent?
    \item[\textbf{Q3}] At what time does the loss caused by the malware threat drop sharply?
    \item[\textbf{Q4}] Do you have an estimation of the proportion of tangible and intangible impacts?
    \item[\textbf{Q5}] Which threat costs the most to address? And which threat is most emergent?
    \item[\textbf{Q6}] Do you have a clear picture of the business impact composition caused by the insider threat?
    \item[\textbf{Q7}] How do participants rate the usability?
\end{enumerate}
\end{quote}

Aside from testing the effectiveness of the features provided by the prototype, open feedback and suggestions on the prototype were collected. Another question related to the usability of the prototype was the collection of the SUS score by means of the ten questions as defined in~\cite{brooke1996sus}. Although SUS presents several limitations (\eg being non-diagnostic and subjective), it is a valuable complement to understanding the perceived subjective view on usability.

\subsubsection{Execution}
Participants were recruited based on a questionnaire to understand their background, education level, and whether they already hold knowledge of cybersecurity management and business impact analysis. Knowledge of cybersecurity was mandatory, although no working proficiency was required. After evaluating 14 participants, the five selected ones shown in Table~\ref{tab:participants} were involved in the experiment since they met those requirements. In addition, a moderator was present to guide the tasks and discussion. As described, only the participant with a finance background had heard of BIA during a course on business continuity management. Furthermore, only two participants had previous experience testing applications.

\begin{table}[H]
\centering
\caption{Participants Involved in the Focus Group Experiment}
\label{tab:participants}
\begin{tabular}{@{}llcc@{}}
\toprule    
           \# & \textit{Background} & \textit{BIA Knowledge} & \textit{Testing Experience}
            \\ \midrule
           1  & Master Student & No & Yes\\
           2  & Software Engineer & No & Yes \\
           3  & Master Student & No & No \\
           4  & Master Student & No & No \\
           5  & Finance Expert & Yes & No \\
           \bottomrule
\end{tabular}
\end{table}

The focus group discussion was held online, lasting 57 minutes. To understand the participants' experimental execution, all screens were recorded. The tasks were executed in turn, while the interaction with the platform was recorded. The methodology and its prototype were introduced in a narrative manner, without giving a live demonstration, which may leak hints on how to use the platform and thereby make it difficult to understand the usability limitations of the platform. Then, the moderator presented the hypothetical scenario and distributed the tasks to the participants. The participants then worked on the tasks independently, with the only support provided being the scenario representation in terms of data provided. While working on the tasks, participants were encouraged to carefully analyze and use each component supplied by the tool to ensure they had a complete view of the tool. Any questions and comments raised during the question that did not directly relate to the scenario were recorded for interpretation. After 35 minutes, all participants had concluded the task, followed by moderated questions to understand the effectiveness of the application and questionnaire distribution to assess the usability.

\subsubsection{Result Analysis}
The analysis of results is based on \1 information collected and recorded during execution (\ie notes and transcripts) and \2 the statically structured questionnaire. Revisiting the initial evaluation questions defined (\cf Section~\ref{sec:focusgroup-methodology}), it can be observed that the results obtained from the execution phase are congruent with the expected results. Again, it has to be stated that there is no ground-truth model. However, it can be analyzed how similar the participants' results are within the group and with the pre-defined scenario. For the first six functional questions, participants came up with congruent results. For \textbf{Q5}, which required understanding not only the most impactful but also the most emergent threat (\ie requiring a timely response), one participant misinterpreted the graphs and gave the inverted result (\ie the least emergent threat with the longest tolerable period). This indicates that the application positively influences the discovery of impact factors behind technical threats, even for users with limited knowledge of business continuity management, given that they have access to estimates or data representing the business perspective. In summary, the answers to questions \textit{Q1} and \textit{Q6} indicate that the participants understood the relevant impact factors for the selected case. Furthermore, the answers to \textit{Q2} to \textit{Q5} demonstrate that participants successfully analyzed and interpreted the threats, leading to the identification of the relevant factors for the three threats (\ie malware attack, botnet, insider threat) and analysis of the quantification by means of impact and emergency.

The usability is investigated by analyzing the SUS questionnaire and interpreting the unstructured data, which comprises the moderator's notes, audio transcriptions, observations, and opinions from the questionnaire. All textual data is gathered, coded, and tagged with a code to simplify content analysis, revealing patterns in user feedback. As revealed by the data, five areas are essential. The first two areas relate to using actionable user interface elements such as buttons and input forms. For example, users were not completely clear on the actions performed by the elements and how they had to structure their input. Thus, additional guidance could improve the guidance in the tool. This is backed by the third and fourth areas, where users expressed that additional textual hints could help explain specific keywords and instructions. This appears to be an essential aspect, as the tool should be usable by users who do not have expertise in the proposed metrics. However, based on the task execution, it can be inferred that although users do not feel confident in certain aspects, they can still use the tool. Finally, users were unclear about how the data collected by the tool is stored, which could be improved by additional notes in the \textit{Home page}. Here, it is crucial to consider that users only received minimal training and that the application only uses client-side storage, which could alleviate this potentially privacy-related concern.

\begin{table}[t]
\centering
\caption{Normalized SUS Scores}
\label{tab:sus}
\begin{tabular}{@{}l|cccccc@{}}
\toprule    
           \textit{Participant} & \textit{1} & \textit{2} & \textit{3} & \textit{4} & \textit{5}
            \\ \midrule
           Score  & 75 & 85 & 55 & 72.5 & 72.5\\
           \bottomrule
\end{tabular}
\end{table}

After identifying areas of concern, the SUS score can be cross-referenced to understand the perceived usability. Combining the ten questions, the SUS scores shown in Table~\ref{tab:sus} can be obtained after normalization. Four out of five participants think the usability is sufficient, while one deems it excellent -- overall, the usability level is acceptable. The SUS score has to be interpreted carefully since it is said to be non-diagnostic. Nevertheless, looking into the distinct categories, the learnability of the application presents \textit{good} results, while the time required to use the tool received inconclusive results, which could indicate a deficit or that the participant group does not precisely reflect the target group at hand.

\subsection{Expert-based Within-Subject Experiment}\label{sec:comparison}
In the final experiment, two widely deployed techniques from the risk assessment and threat analysis field were employed. Based on the results obtained, it is illustrated how the methodology and the results differ, enabling the discussion and illustration of how \sol{} provides additional guidance and a more granular result than existing approaches.

\subsubsection{Risk Matrices}
In the first of the two expert-based exercises, a qualitative two-dimensional (\ie incorporating an estimate of the probability and severity of a threat) risk matrix using three levels (\ie low, medium, high) is employed. The subject present in the experiment holds a Master of Science in Data Science and has more than 7 years of working experience in the computer science industry, working for both enterprises and consultancies in the role of a network engineer, data scientist, and project manager. After working on projects related to implementing a secure development process, the participant considers himself knowledgeable in cybersecurity, although the participant never held a position that exclusively focused on cybersecurity (\eg threat analyst, CISO). This lack of a dedicated security-related position poses both a limitation and an opportunity since the prioritization may lack security expertise. However, this subject is qualified to investigate how well-suited the method is at guiding a non-security expert, which is highly relevant in scenarios such as the Secure Software Development Lifecycle (SSDLC), where software and infrastructure engineers might be tasked with implementing security aspects.

\begin{figure}[t]
    \centering
    \includegraphics[width=.7\linewidth]{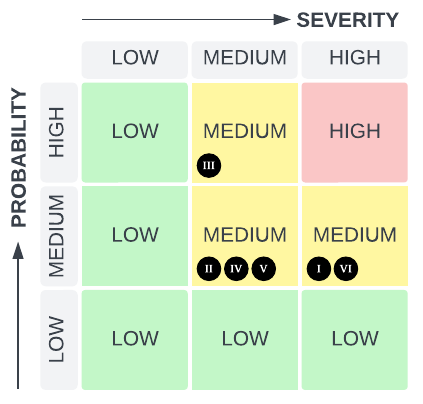}
    \caption{Threat Analysis using a 3x3 Risk Matrix}
    \label{fig:riskmatrix}
\end{figure}

After presenting the scenario obtained from the case study (\cf Section~\ref{sec:casestudy}), additional technical details about the service to be adopted were introduced, and access to the system's technical documentation was granted. Then, the participant was asked to create a prioritization of the threat model defined in Table~\ref{table:threatmodel}, revealing only the threat events. After half an hour, the participant presented the prioritization results, which are noted and visualized in Figure~\ref{fig:riskmatrix}. Comparing the results with the output of \sol{} presented in Table~\ref{table:quantm} demonstrates apparent differences between the two. It has to be restated again that, due to the lack of data for this case, it is not fair to reason on the correctness of either model. However, four observations are apparent. 

The analysis using the matrix \1 demonstrates that all threats are ranked the same, centering around the \textit{medium} score. While it cannot be proven, it is improbable that all threats share the same priority. Apparently, \2 interpreting the results of the matrix is limited; it is not clear whether two threats ranked as \textit{high} lead to the same damage. For example, the DDoS and Ransomware threats are both given a \textit{high} severity, although based on the business impact analysis, one might lead to a few thousand Swiss Francs of direct financial impact. At the same time, the other includes intangible factors, such as the degraded customer relationship due to pricing policies being revealed -- leading to long-term financial losses that may be disastrous. It might be questionable whether a business-critical threat with a \textit{medium} likelihood is not a key concern. 

It is also clear that \3 this qualitative approach presents little diagnostic value since it requires additional interpretation of why a score was achieved and how remediation can be applied. Here, the business impact factors leveraged by \sol{} can highlight weak spots in the company's resilience. For example, in the ransomware example, additional technical measures could be taken to decrease the likelihood of a successful attack. However, from a business continuity perspective, it is clear that the company's posture is not resilient, even in the unlikely case of an attack. Thus, non-technical measures, such as refining pricing policies and contracts, can be considered. Finally, \4 relying on qualitative, subjective estimates for measuring probability may often be favored by professionals due to the lack of data. However, as shown in the case study, data is available for specific threats. Thus, modeling using actual probabilities may be more appropriate, even when some probabilities are estimated or based on industry-wide data (\eg from reports and public datasets).

\subsubsection{DREAD Methodology}
Next, as part of the evaluation, the expert was tasked to create a threat prioritization using the widely known \textit{DREAD} methodology. Initially, said methodology has been proposed (and in the meantime abandoned) by Microsoft. It remains to be widely used. To apply the methodology, a threat is assessed by estimating its impact along five dimensions (\ie Damage, Reproducibility, Exploitability, Affected Users, Discoverability) ranked from zero to ten. The categories carry pre-defined statements to gauge the exposure of a threat to these dimensions. For example, exposure to damage could match the description of \textit{Non-sensitive administrative data being compromised} or \textit{Destruction of an information system; data or application unavailability}. The first example yields a damage score of 9, while the latter yields a maximum score of 10~\cite{dread}. In that sense, the method could be considered opinionated since, in this example, a DDoS attack is assumed to be more damaging than a data breach. Finally, the scores are added to achieve a final scoring that is qualified in four categories (\ie low, medium, high, critical).

\setlength{\tabcolsep}{6pt}

\begin{table}[h]
\centering
\caption{Threat Analysis using DREAD}
\label{tab:dread}
\begin{tabular}{@{}l|ccccc|lr@{}}
\toprule    
           \textit{Threat} & \textit{D} & \textit{R} & \textit{E} & \textit{A} & \textit{D} & \textit{Sum} & \textit{Grade}
            \\ \midrule
           DDoS       & 10   & 10   & 10  & 10   & 8   & 48 &C  \\
           CSRF       & 8   & 0-10  & 5   & 6    & 0-5 & 19-34 & M-H  \\
           XSS        & 8   & 0-7.5 & 5   & 6    & 0-5 & 19-31.5 & M-H  \\
           XXE        & 5   & 5     & 5   & 0-10 & 5   & 20-30& M-H \\
           Deserial.  & 5   & 5     & 5   & 0-10 & 0-10 & 15-35& M-H  \\
           Ransomw.   & 10   & 5    & 2.5   & 10    & 0-10 & 27.5-37.5 & M-H  \\
           \bottomrule
\end{tabular}
L=Low, M=Medium, H=High, C=Critical
\end{table}

After tasking the expert with applying the DREAD methodology by giving access to material on how to apply the methodology and allowing any questions on the aforementioned threat model, the scores are collected for each threat event. As shown in \tablename~\ref{tab:dread}, the expert had difficulties settling for a specific score; hence, the ranges that the expert deems fitting were recorded. The final score was computed by computing the sum of all lower scores and one representing the upper bound. As highlighted in the last column, almost all scores share the same qualification. The only exception here marks the \textit{DDoS} attack, which is assigned a critical rating due to its effect on service availability. While it cannot be disproven as a critical threat, the previously obtained results from the \sol{} approach show a different view. Here, considering the business case's procedural and organizational aspects does not justify criticality. This underlines that the DREAD methodology could be interpreted as being opinionated, which may not match the business in question.

Furthermore, assigning threats into similar categories may indicate that the scoring system is not granular enough to capture the actual impact of threats. For example, prioritizing these threats would lead to focusing on the DDoS attack. It is unclear how to proceed after the exposure to this threat is reduced. Moreover, considering the scores outside of a pure prioritization scenario, it becomes clear that the scores are hard to interpret. For example, the previous quantification using \sol{} discussed whether the DDoS threat's economic impact justifies costly controls. Using these scores may become even more difficult since the numbers do not represent more than the qualitative categories (\ie low, medium, high, critical). In contrast, \sol{} provides specific business impact factors, which incorporate non-tangible impacts.



\subsection{Discussion and Limitations}
Evaluating threat models is a challenging endeavor since there is usually no simple notion of a ground truth threat model that perfectly captures complete information (\eg threat actors, their potential events, and the potential implications on a target system). This is challenging, especially when considering the complexity of the systems to be protected and the security controls governing them~\cite{oppliger}. 
Nevertheless, advancing the field of threat modeling requires a discussion of the strengths and limitations, which are based on the results of our evaluations. The key findings are described as follows.
\begin{enumerate}
    \item Using the quantification model and methodology provided by \sol{} enables precise quantification of a threat's impact. As demonstrated in the case study, this approach is feasible, leading to the discovery of critical threats. As highlighted in the comparison with other approaches, this output is more granular. For example, it enables the distinction between highly impactful and disastrous cases.

    \item Following a quantitative output, related economic approaches from the cybersecurity domain can be leveraged. For example, a quantitative impact can be offset against a security control's cost, revealing its economic efficiency. In practice, it is assumed that this is more feasible for reactive measures (\eg adopting a protection service) than preventative measures (\eg improving secure development practices).

    \item \sol{} is capturing not simply technical complexity but instead is oriented towards the business impact, preserving the semantics of a threat.

    \item Aligned with the previous aspect, \sol{} is not opinionated from a particular technical perspective. As demonstrated in the expert-based experiments, \sol{} disputes the quantification of a threat based on technical terms. For example, using \sol{} a DDoS attack on two different business processes (or enterprises) would yield different impact ratings.

    \item The prototype implementing \sol{} is guided and diagnostic. The focus group evaluation demonstrated that participants with basic cybersecurity familiarity can leverage the guiding prototype to conduct a business impact analysis and discover impact factors. Having such impact factors allows for understanding a threat's context, enabling the adoption of technical and non-technical mitigations.
\end{enumerate}

In that sense, \sol{} addresses relevant limitations in cybersecurity risk management. Nevertheless, it is vital to enumerate the limitations of the performed work. Due to the lack of ground-truth data, assessing the correctness of the model is infeasible. Nevertheless, the case study demonstrated that \sol{} leads to identifying a "disaster threat," clearly distinguished from other relevant threats. Based on the experiences drawn in the evaluations, it appears that the impact assessment is the most value-adding factor to be considered, while others, such as the degree of impact, show less confidence. Reflecting on the overall research methodology adopted, the qualitative nature and selection of a limited number of cases are limiting factors. Nevertheless, this enabled to study the problem in a realistic setting, which is typical for case-study research~\cite{casestudyresearchmethodology}.
\section{Conclusions and Future Work}
\label{conc}
This paper introduced the \sol{} approach, which is composed of a methodology and a quantitative model to assess the impact of threats by incorporating the operational and strategic views from the business side into the threat model. \sol{} defines how to gather the input variables for the quantitative analysis model across all three layers of an organization. The technical threat model is created on the information systems level. On the business process layer, the technical threats are interpreted and translated into business disruptors of a business process. Finally, on the organizational level, the business impact of the disruptors is assessed in terms of tangible and intangible costs, thus, allowing to capture broader consequences of a technical threat.

To assess the feasibility and effectiveness of \sol, the method was applied in a practical case study, where a threat model was created for an information system in a Swiss SME. Based on the experiences drawn from the case study, \sol{} enables quantitative threat prioritization, the formulation of threats for different audiences in an organization, and the discussion on the cost-effectiveness of security controls. Furthermore, the prototype implementing the approach was discussed by means of a focus group experiment, highlighting that non-security practitioners can successfully apply the prototype and its implicit approach, attesting to the perceived usability. Finally, two expert-based experiments have been conducted to contrast the results obtained with \sol{} against two popular threat analysis techniques. Although the results must be cautiously interpreted, the contrasting results may indicate that threat prioritization driven by the semantics of business continuity presents additional benefits, such as increased explainability, more granular statements, and improved usability. Therefore, it can be summarized that a quantitative and business-centric threat modeling approach can help make system engineering activities more systematic, holistic, diverse, and explainable for different stakeholders.

As future work, the \sol{} approach can be evaluated in different scenarios, considering different organization settings. Furthermore, it will be evaluated how a quantitative method can improve development settings where no direct business involvement is available. Finally, a tool incorporating and guiding the \sol{} approach and model can be implemented to make it accessible.  

\section*{Acknowledgments}
This work has been partially supported by \textit{(a)} the Swiss Federal Office for Defense Procurement (armasuisse) with the CyberForce project (CYD-C-2020003) and \textit{(b)} the University of Zürich UZH.

\newpage
\section*{Biography}

\vskip -7\baselineskip

\begin{IEEEbiography}[{\includegraphics
[width=1in,height=1.25in,clip,keepaspectratio]{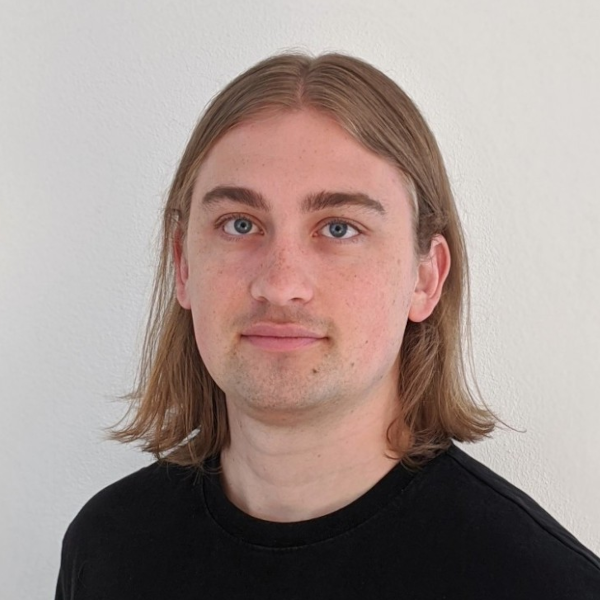}}]
{Jan von der Assen} 
Jan von der Assen received his MSc degree in Informatics from the  University of Zurich. Currently, he is pursuing his Doctoral Degree under the supervision of Prof. Dr. Burkhard Stiller at the Communication Systems Group, University of Zurich. His research interest lies at the intersection between risk management and the mitigation of cyber threats. In his research, he focuses on a holistic perspective on security management, delving into multifaceted aspects such as threat intelligence, vulnerability assessment, and incident response. In addition to his core research focus, his research interests encompass areas such as data privacy, network security, and the integration of emerging technologies in cybersecurity.
\end{IEEEbiography}

\vskip -6\baselineskip

\begin{IEEEbiography}[{\includegraphics
[width=1in,height=1.25in,clip,keepaspectratio]{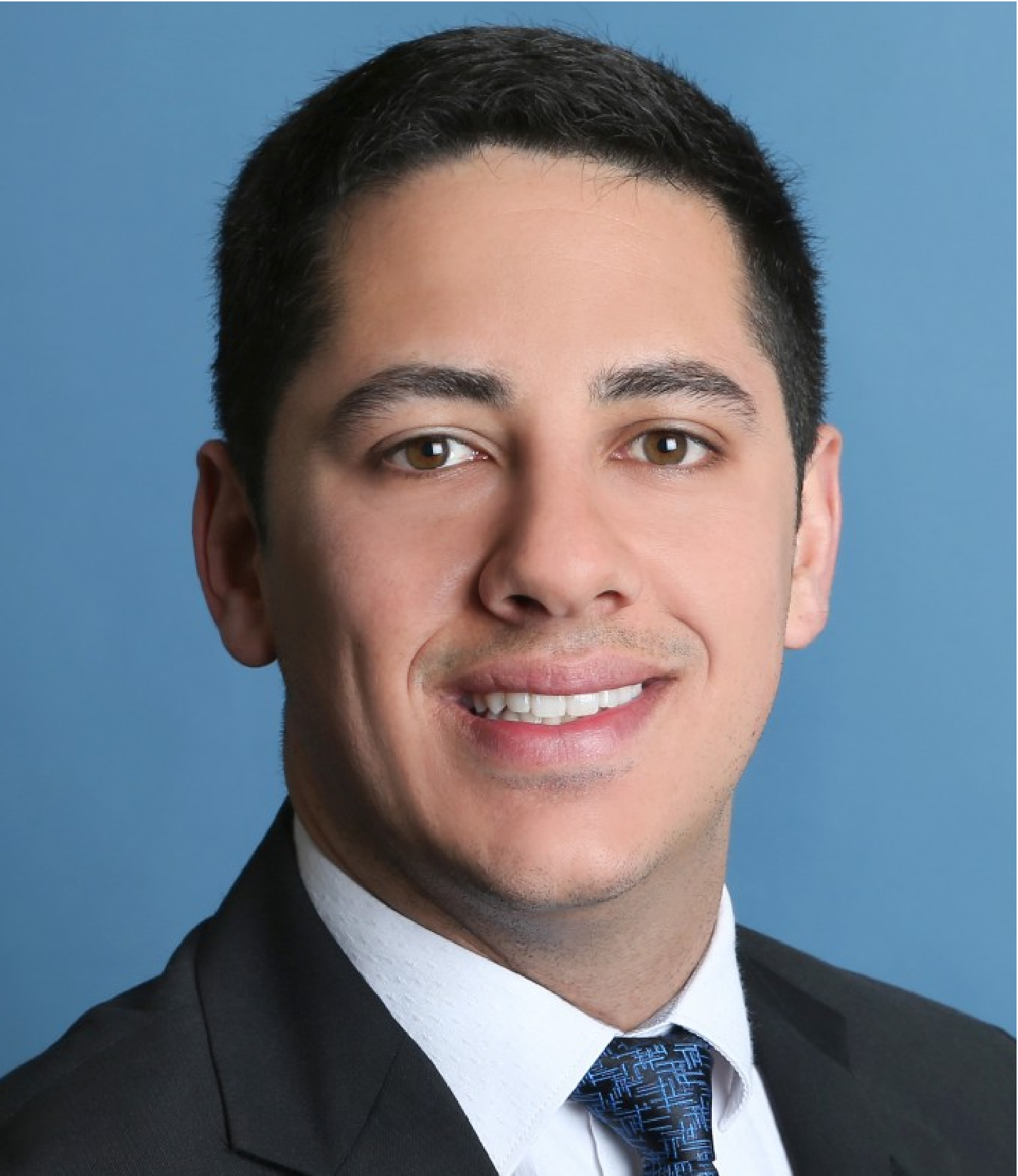}}]
{Muriel F. Franco} 
Dr. Muriel F. Franco is a Guest Researcher affiliated with the University of Z{\"u}rich UZH, Switzerland, within the Communication Systems Group (CSG) of the Department of Informatics (IfI) and a Postdoctoral Researcher at the Computer Networks Group of Federal University of the Rio Grande do Sul (UFRGS). In his research career, he participated in different multidisciplinary projects within teams of networking, cybersecurity, and economic experts, with over 60 research papers and patents co-authored. Muriel holds a PhD (Summa cum laude) from 2023 in Computer Science from the UZH, Switzerland, MBA from 2022 in Project Management from the University of São Paulo (USP), Brazil, MSc from 2017 in Computer Science from UFRGS, Brazil, and received a BSc from 2014 in Computer Science from the Federal University of Pelotas (UFPEL), Brazil. His research interests include cybersecurity, network management, information visualization, and communication systems. 
\end{IEEEbiography}

\vskip -6\baselineskip

\begin{IEEEbiography}[{\includegraphics
[width=1in,height=1.25in,clip,keepaspectratio]{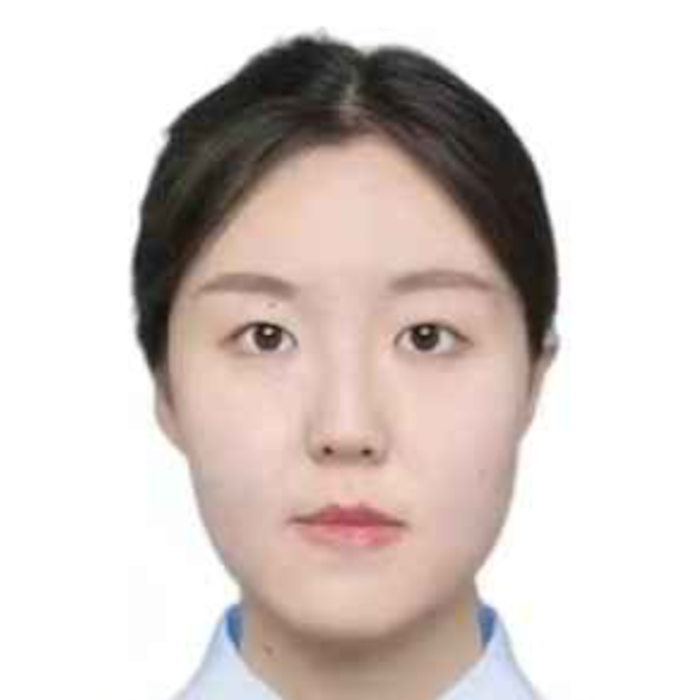}}]
{Muyao Dong} 
Muyao Dong has successfully completed her Master of Science in Data Science from the University of Zurich, building upon her foundational academic background with a Bachelor of Engineering in Computer Science and Technology from the Beijing Institute of Technology. During her academic journey, she dedicated her efforts to a comprehensive exploration of data science principles and methodologies.
Her  academic endeavor took place within the esteemed Communication Systems Group, where she conducted her master's thesis under the guidance of Jan von Der Assen and Prof. Dr. Burkhard Stiller. The focal point of her thesis was an in-depth investigation into cyber threat modeling coupled with a thorough business impact analysis. This research underscored her commitment to advancing the understanding of the intricate interplay between cybersecurity and business resilience.\end{IEEEbiography}
\vskip -6\baselineskip

\begin{IEEEbiography}[{\includegraphics
[width=1in,height=1.25in,clip,keepaspectratio]{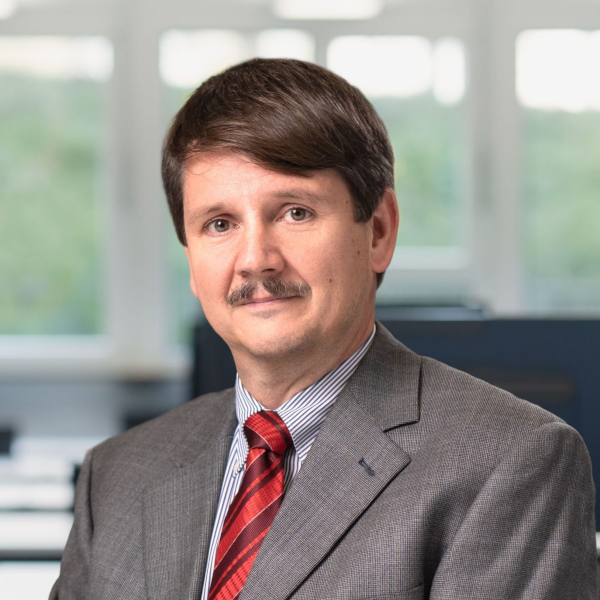}}]
{Burkhard Stiller} 
Burkhard is the director of the Communication Systems Group CSG and Full Professor of Computer Science, in particular on Communication Systems, in the Department of Informatics IfI at the University of Zurich UZH. Burkhard is a member of the UZH Blockchain Center (BCC), a member of the the Center for Information Technology, Society, and Law (ITSL), a member of the Luxembourg Science Foundation’s (FNR) Scientific Council, the Chair of IFIP’s Technical Committee TC6 on Communication Systems, an Editorial Board member of the IEEE Transactions on Network and Service Management (TNSM), the Springer Economic Research and Electronic Networking Series as well as the Journal of Network and Systems Management, Wiley’s International Journal of Network Management, and past Editor-in-Chief of Elsevier’s Computer Networks Journal as well as past chair of the IEEE Computer Society’s Technical Committee on Communication Systems (TCCC).

Burkhard’s research interests include systems with fully decentralized control (blockchains, clouds, peer-to-peer, eVoting), network and service management (economic and security management), Internet-of-Things (security of constrained devices, LoRa), and telecommunication economics (charging and accounting).
\end{IEEEbiography}


\balance

\bibliographystyle{IEEEtranS}
\bibliography{bib/biblio.bib}

\end{document}